\documentclass[prd,aps,showpacs,nofootinbib]{revtex4}
\usepackage{amssymb}
\usepackage{graphicx}
\usepackage{amsmath}

\newcommand{\be}{\begin{equation}}
\newcommand{\ee}{\end{equation}}
\newcommand{\bq}{\begin{eqnarray}}
\newcommand{\eq}{\end{eqnarray}}

\begin{document}

\title{N=1 Supersymetric Quantum Mechanics in a Scenario with
Lorentz-Symmetry Violation}
\author{H. Belich $^{a,e}$, T. Costa-Soares$^{d,e}$, J.A. Helay\"{e}l-Neto$%
^{c,e}$, M.T.D. Orlando$^{a,e}$ and R.C. Paschoal$^{b,e}$}
\affiliation{$^{a}${\small {Universidade Federal do Esp\'{\i}rito Santo (UFES),
Departamento de F\'{\i}sica e Qu\'{\i}mica, Av. Fernando Ferrari, S/N
Goiabeiras, Vit\'{o}ria - ES, 29060-900 - Brasil}}}
\affiliation{$^{b}${\small {Centro Federal de Educa\c{c}\~ao Tecnol\'ogica Celso Suckow
da Fonseca (CEFET/RJ), Av. Maracan\~{a}, 229, 20271-110 - Rio de Janeiro -
RJ - Brasil }}}
\affiliation{{\small {~}}$^{c}${\small {CBPF - Centro Brasileiro de Pesquisas F\'{\i}%
sicas, Rua Xavier Sigaud, 150, CEP 22290-180, Rio de Janeiro, RJ, Brasil}}}
\affiliation{$^{d}${\small {Universidade Federal de Juiz de Fora (UFJF), Col\'{e}gio T%
\'{e}cnico Universit\'{a}rio, Av. Bernardo Mascarenhas, 1283, Bairro F\'{a}%
brica - Juiz de Fora - MG, 36080-001 - Brasil}}}
\affiliation{$^{e}${\small {Grupo de F\'{\i}sica Te\'{o}rica Jos\'{e} Leite Lopes, C.P.
91933, CEP 25685-970, Petr\'{o}polis, RJ, Brasil}}}
\email{belich@cce.ufes.br, tcsoares@cbpf.br, helayel@cbpf.br,
orlando@cce.ufes.br, paschoal@cbpf.br}

\begin{abstract}
We show in this paper that the dynamics of a non-relativistic particle with
spin, coupled to an external electromagnetic field and to a background that
breaks Lorentz symmetry, is naturally endowed with an N=1-supersymmetry.
This result is achieved in a superspace approach where the particle
coordinates and the spin degrees of freedom are components of the same
supermultiplet.
\end{abstract}

\pacs{11.30.Cp, 11.30.Pb,12.60.Jv.}
\maketitle

Lorentz and CPT invariances are cornerstones in modern Quantum Field Theory,
both symmetries being respected by the Standard Model for Particle Physics.
Nevertheless, nowadays one faces the possibility that this scenario is only
an effective theoretical description of a low-energy regime, an assumption
that leads to the idea that these fundamental symmetries could be violated
when one deals with energies close to the Planck scale. Since the pioneering
work by Carroll-Field-Jackiw \cite{Jackiw}, Lorentz-violating theories have
been extensively studied and used as an effective probe to test the limits
of Lorentz covariance. Nowadays, these theories are encompassed
in the framework of the Extended Standard Model (SME), conceived by Colladay
and Kostelecky \cite{Colladay}\ as a possible extension of the minimal
Standard Model of the fundamental interactions. The SME admits Lorentz and
CPT violation in all sectors of interactions by incorporating tensor terms
(generated possibly as vacuum expectation values of a more fundamental
theory) that account for such a breaking. Actually, the SME model sets out
as an effective model that keeps unaffected the $SU(3)\times SU(2)\times U(1)
$ gauge structure of the underlying fundamental theory while it breaks
Lorentz symmetry at the particle frame. 

The gauge sector of the SME has been studied with focus on many different
aspects \cite{Colladay}-\cite{Belich}. The fermion sector has been
investigated as well, initially by considering general features (dispersion
relations, plane-wave solutions, and energy eigenvalues) \cite{Colladay},
and later by scrutinizing CTP-violating probing experiments conceived to
find out in which extent the Lorentz violation may turn out manifest and to
set up upper bounds on the breaking parameters. The CPT theorem, valid for
any local Quantum Field Theory, predicts the equality of some quantities
(life-time, mass, gyromagnetic ratio, charge-to-mass ratio) for a particle
and its corresponding anti-particle. Thus, in the context of Quantum
Electrodynamics, the most precise and sensitive tests of Lorentz and CPT
invariance refer to comparative measurements of these quantities. For the
sake of illustration, a well-known example of this kind of test involves
high-precision measurement of the gyromagnetic ratio \cite{Gyro} and
cyclotron frequencies \cite{Cyclotron} for electron and positron confined in
a Penning trap for a long time. The unsuitability of the\ usual
figure of merit adopted in these works, based on the difference of the
g-factor for electron and positron, was shown in Refs.~\cite{Penning}, in
which an alternative figure of merit was proposed, able to constrain the
Lorentz-violating coefficients (in the electron-positron sector) to $1$\
part in $10^{20}$. Other interesting and precise experiments, also devised
to establish stringent bounds on Lorentz violation, proposed new figures of
merit involving the analysis of the hyperfine structure of the muonium
ground state \cite{Muon}, clock-comparison experiments \cite{Clock},
hyperfine spectroscopy of hydrogen and anti-hydrogen \cite{Hydrog}, and
experiments with macroscopic samples of spin-polarized matter \cite{Spin}.

The influence of Lorentz-violating and CPT-odd terms specifically on the
Dirac equation has been studied in Refs.~\cite{Halmi}, with the evaluation
of the nonrelativistic Hamiltonian and the associated energy-level shifts. A
similar investigation searching for deviations on the spectrum of hydrogen
has been recently performed in Ref.~\cite{Fernando}, where the
nonrelativistic Hamiltonian was derived directly from the modified Pauli
equation. Some interesting energy-level shifts, such as a Zeeman-like
splitting, were then reported. These results may also be used to set up
bounds on the Lorentz-violating parameters.

More recently, in the work of Ref.~\cite{ACNminimo}, fermions have been
reassessed and the influence of a non-minimally coupled Lorentz-violating
background present in the Dirac equation has been investigated. It has been
shown that such a coupling, given in the form $\epsilon _{\mu \nu \alpha
\beta }\gamma ^{\mu }v^{\nu }F^{\alpha \beta }$, is able to induce
topological phases (Aharonov-Bohm and Aharonov-Casher \cite{Casher}) on the
wave function of an electron (interacting with the gauge field and in the
presence of the fixed background). Next, in connection with this particular
effect, it has been found out that (non-minimally coupled) particles and
antiparticles may develop opposite Aharonov-Casher (A-C) phases whenever
particular backgrounds are introduced to realize Lorentz-symmetry violation.
This fact, with a suitable experiment, may be used to constrain the
Lorentz-violating parameter \cite{Thales}. This coupling in the Dirac
equation, with special attention to its nonrelativistic regime and possible
implications on the hydrogen spectrum, was recently studied in \cite%
{Manohidro}. In these papers, the analysis of the non-relativistic regime of
the Dirac equation has revealed that topological quantum phases are induced\
whenever the fermion is coupled to the fixed background and the gauge field
in a non-minimal way. More specifically, it has been found out that
a neutral particle acquires a magnetic moment (induced by the background),
which originates the A-C phase subject to the action of an external electric
field \cite{Casher}. It is worthy stressing that the standard
Aharonov-Casher phase is currently interpreted as due to a Lorentz change in
the observer frame. In our proposal, namely, in a situation where Lorentz
symmetry is violated, it rather emerges as a phase whose origin is ascribed
to the presence of a privileged direction in the space-time, set up by the
fixed background. Since in this kind of model Lorentz invariance in the
particle frame is broken, the Aharonov-Casher effect could not any longer be
obtained by a suitable Lorentz change\ in the observer frame.

Different lines for approaching field theories with Lorentz symmetry
violation in connection with supersymmetry have been carried out in the
works of Ref.~\cite{susybelich}. The present work has as its main goal to
examine the effects of a Lorentz-violating background vector, with
non-minimal coupling \cite{ACNminimo} in the context of an N=1
supersymmetric quantum-mechanical model. A good motivation to justify the
study of Supersymmetric Quantum Mechanics in connection with
Lorentz-symmetry violation is the natural way the spin degrees of freedom
appear: spin interactions do not need to be introduced by hand, they rather
come in as a consequence of the fact that the spin coordinates are the
supersymmetric partners of the space coordinates that describe the
particles. This is very clear if one adopts right from the beginning a
superspace approach, with superfields and actions written directly in
superspace. We could essentially state that our present work sets out to
state that the quantum-mechanical description of a particle with spin,
non-minimally coupled to a background vector that implements the breaking of
Lorentz symmetry, and in the presence of an electromagnetic field has an
intrinsic N=1-supersymmetry, the space coordinate and the spin being the
components of the same supermultiplet. We believe this fact may be a
convincing motivation to support the introduction of supersymmetry.

The N=1-supersymetry approach for a particle\ with spin and non-minimally
coupled to an external electromagnetic field is presented in the sequel,
wich is a generalization to three spatial dimensions of the planar theory of
Ref.~\cite{Paschoal}. The superaction $S$ for the non-relativistic particle
coupled as described above is given as below:%
\begin{equation}
S=\frac{iM}{2}\int dtd\theta D\overrightarrow{X}\cdot \overset{\cdot }{%
\overrightarrow{X}}+iq\int dtd\theta D\overrightarrow{X}\cdot 
\overrightarrow{A}\left( \overrightarrow{X}\right) ,
\end{equation}%
where $\overrightarrow{X}^{j}\left( t,\theta \right) $ is the
supercoordinate of the particle,  $\overrightarrow{A}$ is an arbitrary
differentiable vector function of the supercoordinate and $M$ and $q$ are
real parameters.

$\overrightarrow{X}^{j}\left( t,\theta \right) $ can be expanded according
to: 
\begin{equation}
\overrightarrow{X}^{j}\left( t,\theta \right) =x^{j}+i\theta \lambda
^{j}\left( t\right)
\end{equation}%
with $j=\left( 1,2,3\right) $. $x^{j}$ are the three coordinates of the
particle, $\lambda ^{j}$ are their Grassmannian supersymmetric partners and $%
\theta $ the Grassmannian coordinate that parametrizes the superspace $%
\left( t,\theta \right) .$ By following the conventional approach to
supersymmetry in superspace, we can write the supersymmetry covariant
derivative as follows:%
\begin{equation}
D=\partial _{\theta }-i\theta \partial _{t}.
\end{equation}

Using the fact that for Grassmannian coordinates $\int d\theta =\partial
_{\theta }$, we may write down the Lagrangian in terms of the superfield
components; in so doing, the kinetic term reads as below:%
\begin{equation}
\mathcal{L}_{kin}=\frac{M\overset{_{\cdot }}{x}^{2}}{2}+\frac{iM}{2}\lambda 
\dot{\lambda}.
\end{equation}%
Upon a Taylor expansion of the potential superfield in the superaction and
using the fact that $\lambda $ is a Grassmannian coordinate, a more
conventional \ form of the potential Lagrangian is obtained (after splitting
\ the symmetric and skew-symmetric parts of $\ \partial _{j}A_{i}$): 
\begin{equation}
\mathcal{L}_{int}=q\overset{_{\cdot }}{\overrightarrow{x}}\cdot 
\overrightarrow{A}-\frac{iq}{4}\varepsilon _{ijk}\left[ \lambda _{i},\lambda
_{j}\right] B_{k}.
\end{equation}

The non-minimal coupling to a Lorentz-symmetry violating background vector
was proposed and studied in the works of Ref.~\cite{ACNminimo}. Here, we
can formally introduce this non-minimal coupling by redefining the vector
potential as indicated below: 
\begin{eqnarray}
\overrightarrow{\mathcal{A}} &=&\overrightarrow{A}-\frac{g}{q}v^{0}%
\overrightarrow{B}+\frac{g}{q}\overrightarrow{v}\times \overrightarrow{E},
\label{A1} \\
\widetilde{\vec{B}} &=&\nabla \times \overrightarrow{A}-\frac{g}{q}%
v^{0}\nabla \times \overrightarrow{B}+\frac{g}{q}\nabla \times \left( 
\overrightarrow{v}\times \overrightarrow{E}\right) .  \label{A2}
\end{eqnarray}

Another consequence of the non-minimal coupling is to add the term $q\Phi -g%
\vec{v}\cdot \vec{B}$\ to the Hamiltonian or, equivalently, to subtract this
term in the Lagrangian. However, according to the results of Ref.~\cite%
{Paschoal}, in order to insure the realization of an N=1 supersymmetry, the
condition that follows below must be fulfilled:
\begin{equation}
q\Phi =g\vec{v}\cdot \vec{B}.
\end{equation}

For the sake of convenience, we define a magnetic dipole moment variable, $%
\vec{\mu}$:%
\begin{equation*}
\mu _{k}=-\frac{iq}{4}\varepsilon _{ijk}\left[ \lambda _{i},\lambda _{j}%
\right] ;
\end{equation*}%
with that, we are left with the following interaction Lagrangian:%
\begin{eqnarray}
\mathcal{L}_{int} &=&q\overset{_{\cdot }}{\overrightarrow{x}}\cdot \left[ 
\overrightarrow{A}-\frac{g}{q}\left( v^{0}\overrightarrow{B}-\overrightarrow{%
v}\times \overrightarrow{E}\right) \right] +\overrightarrow{\mu }\cdot (%
\overrightarrow{\nabla }\times \overrightarrow{A})+  \label{A3} \\
&&-\frac{g}{q}v^{0}\overrightarrow{\mu }\cdot (\overrightarrow{\nabla }%
\times \overrightarrow{B})+\frac{g}{q}\overrightarrow{\mu }\cdot 
\overrightarrow{\nabla }\times (\overrightarrow{v}\times \overrightarrow{E}),
\end{eqnarray}%
with the complete Lagrangian written as $\mathcal{L}=\mathcal{L}_{kin}+%
\mathcal{L}_{int}$.

The canonical generalized moment associated to the particle coordinate reads
as:%
\begin{equation}
\vec{p}=\frac{\partial \mathcal{L}}{\partial \overset{_{\cdot }}{%
\overrightarrow{x}}}=M\overset{_{\cdot }}{\overrightarrow{x}}+q\vec{A}%
-g\left( v^{0}\overrightarrow{B}-\overrightarrow{v}\times \overrightarrow{E}%
\right) ,
\end{equation}%
where we observe the Aharonov-Casher phase as a consequence of the breaking
of the Lorentz symmetry. This also illustrates that the coupling of the
particle to the background is the responsible for the appearance of the
magnetic dipole moment, $g\overrightarrow{v}$, as pointed out and discussed
in Ref.~\cite{ACNminimo}.

From this formalism, the general Hamiltonian can be obtained after a
Legendre transformation: 
\begin{equation}
H=\frac{1}{2M}\left[ \vec{p}-q\vec{A}+g\left( v^{0}\overrightarrow{B}-%
\overrightarrow{v}\times \overrightarrow{E}\right) \right] ^{2}-\vec{\mu}%
\cdot \widetilde{\vec{B}}+\left( \pi _{i}+\frac{1}{2}\lambda _{i}\right)
\varepsilon _{i},  \label{kkk24}
\end{equation}%
where the $\pi _{i}$ 's are the fermionic generalized momenta 
\begin{equation}
\pi _{i}=\frac{\partial \mathcal{L}}{\partial \dot{\lambda}_{i}}=-\frac{iM}{2%
}\lambda _{i},
\end{equation}%
that correspond to the second class constraints 
\begin{equation}
\chi _{i}=\pi _{i}+\frac{iM}{2}\lambda _{i}=0
\end{equation}%
and the $\varepsilon _{i}$'s are the corresponding Lagrange multipliers.

The consistence condition, $\dot{\chi}_{i}=0,$ along with the Grassmannian
equations of motion, $\dot{\pi}_{i}=-\frac{\partial H}{\partial \lambda _{i}}
$and $\dot{\lambda}_{i}=-\frac{\partial H}{\partial \pi _{i}}$, allow us to
obtain the Lagrange multipliers:
\begin{equation*}
\varepsilon _{i}=-\frac{q}{M}\varepsilon _{ijk}\lambda _{j}\tilde{B}_{k}.
\end{equation*}%
Substituting this in the Hamiltonian, one obtains:
\begin{equation}
H=\frac{1}{2M}\left( \overrightarrow{p}-q\text{ }\mathcal{\vec{A}}\right)
^{2}+\frac{q}{M}\varepsilon _{ijk}\pi _{i}\lambda _{j}\tilde{B}_{k},
\end{equation}%
which is the same one obtained in the work of Ref.~\cite{ACNminimo} by means
of a different procedure.

Now, regarding the quantization procedure, it is necessary to
calculate the Dirac brackets between all the Hamilton variables. Such a
result has already been obtained in Ref.~\cite{teitel}; as long as the
Grassmannian coordinates, $\lambda _{i}$, are concerned, we get that
\begin{equation}
\left\{ \lambda _{i,}\lambda _{j}\right\} _{DB}=-i\delta _{ij},
\end{equation}%
which, after the quantization procedure, turns into the following
anti-commutator relation $(\hbar =1)$:%
\begin{equation}
\left\{ \hat{\lambda}_{i,}\hat{\lambda}_{j}\right\} =\delta _{ij},
\end{equation}%
where the hats stand for an operator.

As we have shown above, the Lagrangian that contains the
elements for the A-C effect in the presence of Lorentz-symmetry violation
can naturally be described in a supersymmetric quantum framework. The role
played by the spin coordinates, which can be associated to the partners of
the space coordinates, $\lambda _{i}$, is crucial in ensuring the
supersymmetric structure of the model, and it is the Poisson Bracket between 
${\lambda _{i}}$ and ${\lambda _{j}}$, the key ingredient to associating the 
$\lambda $- coordinate components of a spin-1/2 operator, which can be
realized in terms of the Pauli matrices.

So, to conclude, we would like to stress \ and comment on a few issues. By
adopting superspace and superfields, we have been able to present an action
in terms of components where a particle with spin undergoes the
Aharonov-Casher effect. This phase is a natural result of the interaction
between the particle and a background vector that introduces a spatial
anisotropy. Here, we should make a remark concerning the charge of the test
particle. Usually, the A-C effect is presented as an effect of external
electric fields on neutral particles with spin. In our model, we consider,
however, from the beginning, a charged particle. Of course, charged
particles may also develop an A-C phase, but the intrinsic interest of the
effect is for neutral test particles. To describe the latter, we go back to
the interaction Lagrangian of Eq.~(\ref{A3}) and set to zero the parameter $q
$. In so doing, there remain the terms proportional to the components of the
background vector. 

An issue that might deserve some attention regards the relationship
between possible extended supersymmetries and the tensor nature of the
background that implements the violation of Lorentz symmetry. It may happen
that, by introducing the breaking by means of higher rank tensors, the
underlying supersymmetry may be larger than N=1. We know, from the study of
Dirac's equation in the presence of magnetic fields, that, depending on the
field configurations, we may have an N $>$ 1 SUSY. In our case, the
possibility of extended SUSY is connected to the particular tensor character
of the background. This is another point of possible interesting.

As a final comment, we would like to stress that, contrary to what happens
in the case of field theory, the violation of Lorentz symmetry by the
external background vector does not imply the automatic breaking of SUSY.
For fields, the SUSY algebra is based on the assumption of an underlying
Lorentz symmetry; here, in the case of non-relativistic Quantum Mechanics,
the background vector components, $v^{0}$ and $\overrightarrow{v}$, act as
mere parameters while SUSY takes place in a one-dimensional world
parametrized by the time coordinate. In the case of field theory, SUSY is
introduced in  space-time, where $v^{0}$ and $\overrightarrow{v}$ are
components of a four-vector.

Acknowledgments

HB, JAHN and MDTO acknowledge CNPq for the invaluable financial help.

\end{document}